\documentstyle[12pt]{article} 

\newlength{\extraspace}
\newlength{\extraspaces}
\newcommand{\bQ}{{\bar{Q}}}
\newcommand{\bF}{{\bar{F}}}
\newcommand{\QF}{{Q_{\! F}}}
\newcommand{\QS}{{Q_{\! S}}}
\newcommand{\PF}{{P_{\! F}}}
\newcommand{\HF}{{H_{\! F}}}
\newcommand{\bQbF}{{\bar{Q}_{\! \bF}}}
\newcommand{\hQF}{{{\hat{Q}}_{\! F}}}
\newcommand{\hbQbF}{{{\hat{\bar{Q}}}_{\! \bF}}}
\newcommand{\bS}{{\bar{S}}}
\newcommand{\bQS}{{\bQ_{\! S}}}

\newcommand{\bQF}{{\bQ_{\! F}}}
\newcommand{\bB}{{\bar{B}}}
\newcommand{\Bk}{{B_k}}
\newcommand{\Vkn}{{V_{k,n}}}
\newcommand{\WSU}{{W_{\! SU}}}
\newcommand{\WSO}{{W_{\! SO}}}
\newcommand{\PbF}{{P_{\! \bF}}}
\newcommand{\bBox}{{\Box^{\ast}}}
\newcommand{\iibBox}{{{\Box\;\!\!\!\Box}^{\ast}}}
\newcommand{\iiBox}{{{\Box\;\!\!\!\Box}}}
\newcommand{\bq}{{\bar{q}}}

\catcode`\@=11
\def\numberbysection{\@addtoreset{equation}{section}
\def\theequation{\arabic{section}.\arabic{equation}}}
\begin{document}
\thispagestyle{empty}
\begin{center}

\begin{flushright}
NHCU-HEP-97-09 \\
%{\tt hep-th/9707} \\
July, 1997 \\
\end{flushright}
\vspace{3mm}

\begin{center}
{\Large
{\bf  A Comment on Duality in SUSY $SU(N)$ Gauge Theory with 
      a Symmetric Tensor 
}} 
\\[18mm]
{\sc Wang-Chang Su}\footnote{
\tt e-mail: suw@phys.nthu.edu.tw} \\[4mm]
{\it Department of Physics, National Tsing-Hua University \\[2mm]
Hsinchu 300, Taiwan} \\[4mm]

\end{center}
\vspace{18mm}
{\bf Abstract}\\[5mm]
{\parbox{13cm}{\hspace{5mm}
We suggest an alternative approach to deconfine ${\cal N} =1$ $SU(N)$
supersymmetric gauge theory with a symmetric tensor, fundamentals, 
anti-fundamentals, and no superpotential. 
It is found that although the dual prescription derived by this new 
method of deconfinement is different from that by the original method, 
both dual prescriptions are connected by duality transformations. 
By deforming the theory, it is shown that both dual theories flow 
properly so that the Seiberg's duality is preserved.

\vspace{2mm}

PACS numbers: $11.15.{\rm -q}, 11.15.{\rm Tk}$, $11.30.{\rm Pb}$

Key words: supersymmetric gauge theory, duality
}}

\end{center}
\vfill
\newpage

%%%%%%%%%%%%%%%%%%%%%%%%%%%%%%%%%%%%%%%%%%%%%%%%%%%%%%%%%%%%%%%%%%%%%%%
%%%%%%%%%%%%%%%%%%%%%%%%%%%%%%%%%%%%%%%%%%%%%%%%%%%%%%%%%%%%%%%%%%%%%%%
%%%%%%%%%%%%%%%%%%%%%%%%%%%%%%%%%%%%%%%%%%%%%%%%%%%%%%%%%%%%%%%%%%%%%%%
%%%%%%%%%%%%%%%%%%%%%%%%%%%%%%%%%%%%%%%%%%%%%%%%%%%%%%%%%%%

The recent years have witnessed significant development in 
understanding strongly coupled dynamics of supersymmetric gauge 
theories \cite{Sei1,IRSei}. 
See \cite{IS;lec} for recent reviews and references therein for earlier 
work. 
One of the most remarkable results is made by N. Seiberg \cite{Sei2} 
stating that different SUSY gauge theories in the ultraviolet can have 
the same interacting superconformal fixed point in the infrared. 
The fixed point theory is in a non-Abelian Coulomb phase that has dual 
prescriptions in terms of any of the ultraviolet theories. 
This non-Abelian duality is the generalization of Montonen-Olive 
electric-magnetic duality \cite{Mon}.

The original example of ${\cal N} = 1$ duality is that an $SU(N)$ 
gauge theory with $F$ quarks $Q_i$ and ${\tilde Q}_i$ in fundamental 
and anti-fundamental representations of the gauge theory is dual to 
an $SU(F - N)$ gauge theory with $F$ dual quarks $q_i$ and 
${\tilde q}_i$ in fundamental and anti-fundamental representations 
and a set of gauge singlet superfields. 
Both theories have the same global symmetries even though the gauge 
groups are different. 
No proof of this duality is known but the evidence of its existence 
has been supported by several nontrivial consistency tests. 
The duality of $SO(N_c)$ gauge group with matter in vector 
representation \cite{IS;so} and $Sp(N_c)$ gauge group with matter 
in fundamental representation \cite{IP;sp} were also investigated 
in details. 

The duality of SUSY gauge theories with a tensor superfield $X$ is 
also studied \cite{KSS} (see \cite{Chi-nonChi} for other types of 
duality.) 
There are two approaches to investigate the problem that are similar 
but differ only in whether a superpotential $W(X)$ is included or not. 
Because the dynamics are simplified when a superpotential is turned on, 
many examples of non-trivial fixed points with dual prescriptions 
have been reported \cite{Decon,IRSt,Prod,Aha}. 
The addition of a superpotential $W(X)$ has the effects of lifting 
the flat direction and truncating the chiral ring of the theory 
without a superpotential. 
It also has an effect of driving the system to a new infrared fixed 
point when the added superpotential $W(X)$ is a relevant operator. 
When it is irrelevant, the added superpotential has in general 
strong effects on the infrared dynamics. 
If the duality of a theory is not known, a superpotential can be 
introduced to analyze the long distance physics of the theory. 

Examples of the duality of gauge theories with a ranked-two tensor 
field $X$ and no superpotential can be found in
\cite{Decon2,Csaki,Sp,SO}. 
The standard method to investigate the duality of these theories 
is the deconfining technique \cite{Decon,Decon2}. 
In \cite{Decon}, it was suggested that a ranked-two tensor field 
can be considered as a bound state of an auxiliary gauge group. 
Since the auxiliary gauge group confines at low energy, the theory 
with the tensor field $X$ has an equivalent description in terms of 
an expanded theory with a product of two gauge groups. 
In particular, the field contents in the expanded theory are all 
in fundamental or anti-fundamental of the product gauge groups. 
By applying Seiberg's duality, it is found that the dual theory of the 
original theory must be a theory of a product of two gauge groups. 

In this letter, we construct an alternative approach to deconfine 
the $SU(N)$ gauge theory with a symmetric tensor field $S$, $F$ quarks 
$\QF$ in fundamental, $\bF$ quarks $\bQbF$ in anti-fundamental, and 
no superpotential. 
This model, which will be called the electric theory hereafter, has 
been studied by T. Sakai \cite{SO} using the original technique of 
deconfining in which the symmetric tensor field $S$ of the electric 
theory is thought of as a bound state of an auxiliary gauge theory. 
In our new approach, the symmetric tensor field $S$ and the $F$ 
fundamentals $\QF$ of the electric theory are both considered as 
bound states of another auxiliary gauge group. 
Nevertheless, it is shown that these two approaches of deconfining 
reproduce the same electric theory at low energy. The duality of the 
electric theory is then derived from the elementary duality of Seiberg. 
It is found that four dual theories all with an 
\( SU(F+5) \times SO(2F+8) \) gauge group construct a diagram of 
duality flow (see Figure 1) in which two of them are the magnetic 
prescriptions of the electric theory. 
There are two types of duality acting among these theories, namely, 
duality A and duality B. 
Duality A is derived by the original method of deconfining and 
connects two theories with different matter contents and 
transformation properties. 
Duality B is obtained by the new approach of deconfining presented 
in this letter. 
This new duality connects two theories that are charge conjugated 
theories of the $SU(F+5)$ gauge group. 
Any two of the dual theories can be related by a combination of 
duality transformations. 
As for a consistency check of duality, the expectation values of 
superfields in the electric theory are turned on in various ways. 
It is found that both magnetic theories are properly deformed in such 
a way that the non-Abelian duality between them is still preserved. 
By deforming the electric theory along a flat direction, it is shown 
that both magnetic theories flow to the Seiberg's duality in $SO(N)$ 
gauge group.

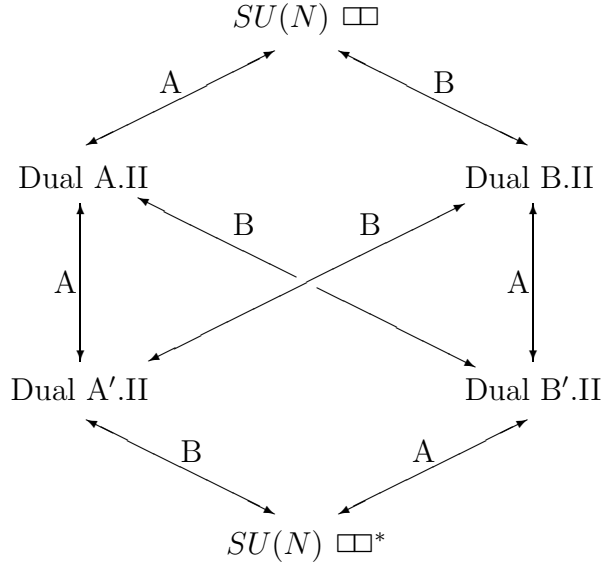
\begin{figure*}[h]
\setlength{\unitlength}{0.7mm}
\begin{picture}(200,110)(-60,0)
  \put(40,0){\makebox(20,10){$SU(N)$ $\iibBox$}}
  \put(0,30){\makebox(20,10)[r]{Dual A$^\prime$.II}}
  \put(0,70){\makebox(20,10)[r]{Dual A.II}}
  \put(40,100){\makebox(20,10){$SU(N)$ $\iiBox$}}
  \put(80,30){\makebox(20,10)[l]{Dual B$^\prime$.II}}
  \put(80,70){\makebox(20,10)[l]{Dual B.II}}
  \put(7,55){\vector(0,1){15}}
  \put(7,55){\vector(0,-1){15}}
  \put(4,53){$\!\! {\rm A}$}
  \put(93,55){\vector(0,1){15}}
  \put(93,55){\vector(0,-1){15}}
  \put(90,53){$\!\! {\rm A}$}
  \put(26,20){\vector(-2,1){17.88}}
  \put(26,20){\vector(2,-1){17.88}}
  \put(28,21){$\!\! {\rm B}$}
  \put(74,20){\vector(2,1){17.88}}
  \put(74,20){\vector(-2,-1){17.88}}
  \put(72,21){$\!\! {\rm A}$}
  \put(26,90){\vector(2,1){17.88}}
  \put(26,90){\vector(-2,-1){17.88}}
  \put(24,91){$\!\! {\rm A}$}
  \put(74,90){\vector(-2,1){17.88}}
  \put(74,90){\vector(2,-1){17.88}}
  \put(76,91){$\!\! {\rm B}$}
  \put(50,55){\vector(2,1){30}}
  \put(48,56){\vector(-2,1){30}}
  \put(38,64){$\!\! {\rm B}$}
  \put(52,54){\vector(2,-1){30}}
  \put(50,55){\vector(-2,-1){30}}
  \put(62,64){$\!\! {\rm B}$}
\end{picture}
\caption{
The duality flow of the $SU(N)$ gauge theory with a symmetric tensor. 
The dual prescriptions of the theory are connected by two types of 
duality. 
Duality A is obtained by expanded method A whereas duality B is 
by expanded method B suggested in this letter.}
\end{figure*}

We start with the electric theory. 
Under the global non-anomalous \( SU(F) \times SU(\bar{F}) 
\times U(1)_1 \times U(1)_2 \times U(1)_R \) symmetries, 
the fields transform as:
\begin{equation}
\label{electric;theory}
\matrix{
& SU(N) & SU(F) & SU(\bF) & U(1)_1 & U(1)_2 & U(1)_R  \cr
S  & \iiBox & {\bf 1} & {\bf 1} & 0& -2F & 0  \cr
\QF  & \Box & \Box & {\bf 1} & 1 & N-F & \frac{2(F+3)}{F}  \cr
\bQbF  &~\bBox & {\bf 1} & \Box & - \frac{F}{\bF} & F & 0
}
\end{equation}
where $\bF = N + F + 4$ is required by the cancellation of $SU(N)^3$ 
gauge anomaly. 
Note that the assignment of the $U(1)_R$ is chosen for convenience. 

The flat directions can be conveniently described by the following 
gauge invariant chiral operators: \( M \equiv \QF \bQbF \), 
\( H \equiv S \bQbF \bQbF \), \( \bar{B} \equiv (\bQbF)^N \), 
\( \Bk \equiv (S)^{N-k} (\QF \QF)^k \) 
\( ( k \le {\rm min} (N,F) ) \), and some operators of exotic 
composites 
\( \Vkn \equiv (\QF)^k (\bQbF)^{N-k-2n} (S)^{N-k-n} (\WSU)^n \) 
(\( k \le {\rm min} (N,F) \) and \( n =0,1,2 \)) contracted with 
one or two $SU(N)$ epsilon tensors. 
$\WSU$ is the field strength superfield of \( SU(N) \) gauge group. 
Not all the gauge invariant operators are independent. 
Some satisfy classical constraints. 
We will not discuss these constraints. 
When $M$ gets an expectation value of rank $r$, the theory is 
higgsed to $SU(N-r)$ with the remaining massless field contents 
being a symmetric tensor $\hat{S}$, $F$ fundamentals $\hQF$, 
and $\bF-r$ anti-fundamentals $\hbQbF$. 
Similarly, when $H$ gets an expectation value of rank $r$, 
the theory is higgsed to $SU(N-r)$ with a symmetric tensor $\hat{S}$, 
$F$ fundamentals $\hQF$, and $\bF-r$ anti-fundamentals 
$\hbQbF$ remained. 
When $\Bk$ gets an expectation, the theory is higgsed to $SO(N-k)$ 
with massless filed contents being $\bF + F-k$ fundamentals $\hat{Q}$. 
When $\bar{B}$ gets an expectation value, the theory is completely 
higgsed. 
In particular, when $\Bk$ gets an expectation value and $H$ gets an 
expectation value of rank $r$, the theory is higgsed to $SO(N-k-r)$ 
with $\bF+F-k-r$ fundamentals remained.

Using holomorphy, symmetry, and weak coupling, there is no 
superpotential that can be generated dynamically for all $F$. 
This leads to the fact that the singularity at the origin of the 
classical moduli space cannot be smoothed out quantum mechanically. 
This singularity cannot be attributed to the gauge invariant 
operators mentioned above neither, because 't Hooft anomaly matching 
conditions are not satisfied.
Consequently, the theory is in non-Abelian Coulomb phase for 
$F \le 2N-3$, which has dual description.

To lay the background for this study, we briefly summarize 
the expanded method of \cite{SO} in our notation. 
It was suggested in \cite{SO} that the electric theory 
(\ref{electric;theory}) has an equivalent expanded theory of a 
product of \( SU(N) \times SO(N+5) \) gauge groups (expanded theory A). 
The field contents and their transformation properties are listed 
below.
\begin{equation}
\label{expanded;theoryA}
\matrix{ 
& SU(N) & SO(N+5) & SU(F) & SU(\bF) & U(1)_1 & U(1)_2 & U(1)_R  \cr
\QS  & \Box & \Box & {\bf 1} & {\bf 1} & 0 & -F & 0  \cr 
P  & {\bf 1} & \Box & {\bf 1} & {\bf 1} & 0 & NF & -2  \cr 
\bQ  &~\bBox & {\bf 1} & {\bf 1} & {\bf 1} & 0 & -F(N-1) & 4  \cr
u  & {\bf 1} & {\bf 1} & {\bf 1} & {\bf 1} & 0 & -2NF & 6  \cr
\QF  & \Box & {\bf 1} & \Box & {\bf 1} & 1 & N-F & \frac{2(F+3)}{F}  
     \cr 
\bQbF  &~\bBox & {\bf 1} & {\bf 1} & \Box & -\frac{F}{\bF} & F & 0 
}
\end{equation}
The expanded theory A has the superpotential
\begin{equation}
W = P^2 u + \QS \bQ P.
\label{superpotential;expandedA}
\end{equation} 

At the scale \( \Lambda_{SO(N+5)} \) \( (\gg \Lambda_{SU(N)}) \), 
the $SO(N+5)$ gauge theory in (\ref{expanded;theoryA}) confines and 
may yield no superpotential since there is a branch on the moduli 
space of vanishing superpotential \cite{IS;so}. 
After massive fields of the confined theory are integrated out, 
the electric theory emerges. 
The duality of the expanded theory A can be constructed by assuming 
holomorphy of the ratio \( \Lambda_{SU(N)} / \Lambda_{SO(N+5)} \). 
When \( \Lambda_{SU(N)} \gg \Lambda_{SO(N+5)} \), 
the expanded theory A has a dual prescription in terms of an 
\( SU(F+5) \times SO(N+5) \) gauge theory (dual A.I theory). 
Now using holomorphy of \( \Lambda_{SU(F+5)} / \Lambda_{SO(N+5)} \), 
the author of \cite{SO} found that the dual A.I theory has a dual 
prescription in terms of a product of \( SU(F+5) \times SO(2F+8) \) 
gauge groups. 
The result of integrating out massive fields is called the first 
magnetic theory or the dual A.II theory which has the following field 
contents:
\begin{equation}
\label{dual2;theoryA}
\matrix{
& SU(F+5) & SO(2F+8) & SU(F) & SU(\bF) & U(1)_1 & U(1)_2 & U(1)_R  
    \cr
\bS  &~\iibBox & {\bf 1} & {\bf 1} & {\bf 1} & {2F \over F+5} & 0 
&{4(F+3) \over F+5}  \cr 
\bQF  &~\bBox & {\bf 1} &~\bBox & {\bf 1} & {-5 \over F+5} & -N 
& -{10(F+3) \over F(F+5)}  \cr
\QS  & \Box & \Box & {\bf 1} & {\bf 1} & {-F \over F+5} & 0
&-{F+1 \over F+5}  \cr
Q  & \Box & {\bf 1} & {\bf 1} & {\bf 1} & {-F \over F+5} & NF 
&-{4(F+4) \over F+5}  \cr
\PbF  & {\bf 1} & \Box & {\bf 1} &~\bBox & {F \over \bF} & 0 & 1  \cr
M  & {\bf 1} & {\bf 1} & \Box & \Box &{N+4 \over \bF} & N 
& {2(F+3) \over F}  \cr 
H  & {\bf 1} & {\bf 1} & {\bf 1} & \iiBox & {-2F \over \bF} & 0 & 0  
      \cr
N  & {\bf 1} & {\bf 1} & \Box & {\bf 1} & 1 & -N(F-1) 
   & {6(F+1) \over F} \cr 
u  & {\bf 1} & {\bf 1} & {\bf 1} & {\bf 1} & 0 & -2NF & 6 
}
\end{equation}
and the superpotential 
\begin{equation}
W = M \bQF \QS \PbF + N Q \bQF + \bS Q^2 u + \bS \QS^2 + H \PbF^2.
\label{superpotential;dual2A}
\end{equation}

Because the $SU(F+5)$ gauge group in the dual A.II theory has field 
contents of a symmetric tensor, it can be expanded by the method of 
deconfining (\ref{expanded;theoryA}) described above. 
The resulting theory is again a dual prescription with an 
\( SU(F+5) \times SO(2F+8) \) gauge theory. 
We refer to this as the dual A$^\prime$.II theory with the following 
matter contents and transformation properties:
\begin{equation}
\label{dual2;theoryAA}
\matrix{ 
& SU(F+5) & SO(2F+8) & SU(F) & SU(\bF)  \cr
S  & \iiBox & {\bf 1} & {\bf 1} & {\bf 1}  \cr
\QF  & \Box & {\bf 1} & \Box & {\bf 1}  \cr
\bQS  &~\bBox & \Box & {\bf 1} & {\bf 1}  \cr
\bQ  &~\bBox & {\bf 1} & {\bf 1} & {\bf 1}  \cr
\PF  & {\bf 1} & \Box &~\bBox & {\bf 1}  \cr
\PbF  & {\bf 1} & \Box & {\bf 1} &~\bBox  \cr 
M  & {\bf 1} & {\bf 1} & \Box & \Box  \cr 
H  & {\bf 1} & {\bf 1} & {\bf 1} & \iiBox  \cr
N  & {\bf 1} & {\bf 1} &~\bBox & {\bf 1}  \cr 
u  & {\bf 1} & {\bf 1} & {\bf 1} & {\bf 1}
}
\end{equation}
The dual A$^\prime$.II theory has the superpotential of this form
\begin{eqnarray}
W = M \PbF \PF + \PF \QS \bQF + N Q \bQF + \bS Q^2 u + \bS \QS^2 
+ H \PbF^2
\label{superpotential;dual2AA}
\end{eqnarray}
Note that the result of dual A$^\prime$.II theory is not a product of 
three gauge groups as we originally expect, because some fields 
acquire masses by Higgs mechanism which breaks the two additionally 
expanded gauge groups \( SO(2F+8) \times SO(2F+8) \) to the diagonal 
subgroup $SO(2F+8)$.
\footnote{
The result can also be obtained without the Higgs mechanism. 
The expanded theory of the dual A.II theory is a product of three 
gauge groups \( SU(F+5) \times SO(F+10) \times SO(2F+8) \). 
In this theory, the $SU(F+5)$ has $2F+10$ flavors and can be dualized 
by the $SU(N)$ duality of \cite{Sei2}, generating another 
\( SU(F+5) \times SO(F+10) \times SO(2F+8) \) gauge theory in which 
some fields acquire masses. 
After integrating out the massive fields, we find that the $SO(F+10)$ 
gauge theory confines. 
As a result, the theory reduces to the dual A.II theory with 
\( SU(F+5) \times SU(2F+8) \) gauge groups.
} 
After integrating out massive fields, we recover the magnetic theory
(\ref{dual2;theoryAA}).

To this end, we suggest an alternative approach to expand the 
electric theory (\ref{electric;theory}). 
Instead of considering the symmetric tensor $S$ as a bound state of 
the $SO(N+5)$ gauge theory as suggested in (\ref{expanded;theoryA}), 
we treat both the symmetric tensor $S$ and the quarks $\QF$ 
in fundamental of the electric theory as bound states of an 
$SO(\bF+1)$ gauge theory. 
This is the expanded theory B with \( N+F+1\) fields in fundamental 
and \( N + \frac{1}{2} (F+1)(F+2) \) singlet fields. 
In order to preserve the same global symmetries as the electric 
theory, a proper superpotential term is added to break the flavor 
symmetry \( SU(N+F+1) \times SU(N+\frac{1}{2}(F+1)(F+2)) \) to 
\( SU(N) \times U(1)^2 \). 
The transformation properties of these fields and the superpotential 
are listed below:
\begin{equation}
\label{expanded;theoryB}
\matrix{ 
& SU(N) & SO(\bF + 1) & SU(F) & SU(\bF) & U(1)_1 & U(1)_2 & U(1)_R  
\cr
\QS  & \Box & \Box & {\bf 1} & {\bf 1} & 0 & -F & 0  \cr 
\PF  & {\bf 1} & \Box & \Box & {\bf 1} & 1 & N & \frac{2(F+3)}{F}  
\cr
P  & {\bf 1} & \Box & {\bf 1} & {\bf 1} & -F & 0 & -2(F+4)  \cr 
\bQ  &~\bBox & {\bf 1} & {\bf 1} & {\bf 1} & F & F & 2(F+5)  \cr
N  & {\bf 1} & {\bf 1} &~\bBox & {\bf 1} & F-1 & -N 
& \frac{2(F^2+4F-3)}{F}  \cr
\HF  & {\bf 1} & {\bf 1} &~\iibBox & {\bf 1} & -2 & -2N & 
\frac{-2(F+6)}{F}  \cr 
u  & {\bf 1} & {\bf 1} & {\bf 1} & {\bf 1} & 2F & 0 & 2(2F+9)  \cr
\bQbF  &~\bBox & {\bf 1} & {\bf 1} & \Box & \frac{-F}{\bF} & F & 0
}
\end{equation}
and
\begin{equation}
W = P^2 u + P N \PF + \HF \PF^2 + \QS \bQ P.
\label{superpotential;expandedB}
\end{equation}

The equivalence of the expanded theory B (\ref{expanded;theoryB}) and 
the electric theory (\ref{electric;theory}) can be understood. 
At the scale \( \Lambda_{SO(\bF+1)} \gg \Lambda_{SU(N)} \) in which 
$SU(N)$ gauge interaction is weakly coupled and can be treated as 
flavor symmetry. 
The $SO(\bF+1)$ gauge theory confines without generating 
superpotential due to the existence of a branch on the moduli space 
of zero superpotential \cite{IS;so}. 
Adding the superpotential (\ref{superpotential;expandedB}) to the 
theory gives masses to all gauge invariant fields of $SO(\bF+1)$ 
except for the fields $S=Q_S^2$ and $\QF = \QS \PF$. 
Now gauge the $SU(N)$ flavor symmetry, under which $S$ is a symmetric 
tensor and $\QF$ is in fundamental, and include an $SO(\bF+1)$ 
singlet $\bQbF$, which is in anti-fundamental of $SU(N)$ gauge 
theory, to cancel the $SU(N)^3$ gauge anomaly. 
The $SU(N)$ $\times$ $SO(\bF+1)$ expanded theory B 
(\ref{expanded;theoryB}) is thus equivalent to the electric theory
(\ref{electric;theory}).

The duality of the expanded theory B (\ref{expanded;theoryB}) can 
then be constructed by assuming holomorphy of the ratio 
\( \Lambda_{SU(N)} / \Lambda_{SO(\bF+1)} \). 
When \( \Lambda_{SU(N)} \gg \Lambda_{SO(\bF+1)} \) the expanded 
theory B is considered as an $SU(N)$ gauge theory with $F+\bF+1$ 
flavors. 
It is in the non-Abelian Coulomb phase that can be dualized by 
the $SU(N)$ duality of \cite{Sei2}. 
The result is a dual prescription in terms of a product of 
\( SU(F+5) \times SO(\bF+1) \) gauge groups. 
With the massive fields being integrated out, the theory becomes
\begin{equation}
\label{dual1;theoryB}
\matrix{ 
& SU(F+5) & SO(\bF+1) & SU(F) & SU(\bF) & U(1)_1 & U(1)_2 & U(1)_R  
\cr 
\bar{q}_S  &~\bBox & \Box & {\bf 1} & {\bf 1} & 0 & \frac{-NF}{F+5}  
           & 0  \cr 
\bar{q}_\bF  &~\bBox & {\bf 1} & {\bf 1} &~\bBox & \frac{F}{\bF} 
& \frac{NF}{F+5} & 2(2F+9)  \cr 
q  & \Box & {\bf 1} & {\bf 1} & {\bf 1} & -F & \frac{FN}{N+5} 
   & -2(F+4) \cr 
p_F  & {\bf 1} & \Box & \Box & {\bf 1} & 1 & N & {2(F+3) \over F}  
\cr 
p_\bF  & {\bf 1} & \Box & {\bf 1} & \Box & \frac{-F}{\bF} & 0 & 0  
\cr
\HF  & {\bf 1} & {\bf 1} &~\iibBox & {\bf 1} & -2 & -2N 
     & {-2(F+6) \over F}  \cr 
N  & {\bf 1} & {\bf 1} &~\bBox & {\bf 1} & F-1 & -N 
   & \frac{2(F^2+4F-3)}{F}  \cr
u  & {\bf 1} & {\bf 1} & {\bf 1} & {\bf 1} & 2F & 0 & 2(2F+9)  \cr 
}
\end{equation}
The superpotential is given by \( W = p_\bF q_\bF \bq_S 
+ N q \bq_S p_F + (q \bq_S)^2 u + \HF p_F^2 \). 
This theory is referred to as the dual B.I theory. It can be checked 
that the 't Hooft anomaly matching conditions are satisfied.

Now use holomorphy for \( \Lambda_{SU(F+5)} / \Lambda_{SO(\bF+1)} \).
Observe that the $SO(\bF+1)$ gauge theory in the dual B.I theory 
(\ref{dual1;theoryB}) has $\bF+2F+5$ flavors in fundamental. 
It is in non-Abelian Coulomb phase and can be dualized by the $SO(N)$ 
duality of \cite{IS;so}. 
After we integrate out the massive fields, the result is again a 
dual prescription of a product of two gauge groups. 
This is the second magnetic theory of the electric theory or 
the dual B.II theory,
\begin{equation}
\label{dual2;theoryB}
\matrix{ 
& SU(F+5) & SO(2F+8) & SU(F) & SU(\bF) & U(1)_1 & U(1)_2 & U(1)_R  
\cr
\bS  &~\iibBox & {\bf 1} & {\bf 1} & {\bf 1} & 0 & \frac{-2NF}{F+5} 
     & 0  \cr
\bQF  &~\bBox & {\bf 1} & \Box & {\bf 1} & 1 & \frac{5N}{F+5} 
& \frac{2(F+3)}{F}  \cr
\QS  & \Box & \Box & {\bf 1} & {\bf 1} & 0 & \frac{NF}{F+5} & 1  \cr
Q  & \Box & {\bf 1} & {\bf 1} & {\bf 1} & -F & \frac{NF}{F+5} 
   & -2(F+4)  \cr
\PF  & {\bf 1} & \Box &~\bBox & {\bf 1} & -1 & -N 
     & \frac{-(F+6)}{F}  \cr 
\PbF  & {\bf 1} & \Box & {\bf 1} &~\bBox & \frac{F}{\bF} & 0 & 1  \cr
M  & {\bf 1} & {\bf 1} & \Box & \Box & \frac{N+4}{\bF} & N 
   & \frac{2(F+3)}{F}  \cr 
H  & {\bf 1} & {\bf 1} & {\bf 1} & \iiBox & \frac{-2F}{\bF} & 0 & 0  
\cr
N  & {\bf 1} & {\bf 1} &~\bBox & {\bf 1} & F-1 & -N
   & \frac{2(F^2+4F-3)}{F}  \cr 
u  & {\bf 1} & {\bf 1} & {\bf 1} & {\bf 1} & 2F & 0 & 2(2F+9)
}
\end{equation}
with the superpotential 
\begin{eqnarray}
W = M \PbF \PF + \PF \QS \bQF + N Q \bQF + 
    \bS Q^2 u + \bS \QS^2 + H \PbF^2.
\label{superpotential;dual2B}
\end{eqnarray}
Note that the 't Hooft anomaly matching conditions are satisfied. 

In the dual B.II theory, its $SU(F+5)$ gauge group has matter 
contents in symmetric tensor, fundamental, and anti-fundamental 
representations, and thus can be dualized by either methods of 
deconfining. 
If the symmetric tensor $\bS$ of the dual B.II theory is deconfined 
by the expanded method A (\ref{expanded;theoryA}), 
the dual B$^\prime$.II theory is obtained. 
This is an \( SU(F+5) \times SO(2F+8) \) gauge theory with field 
contents similar to those of the dual A.II theory except that the 
$SU(F+5)$ gauge fields are given in complex conjugate 
representations, that is, the dual B$^\prime$.II theory and 
dual A.II theory are related by charge conjugation of the 
$SU(F+5)$ gauge group. 
Note that the same phenomenon also occurs between the dual B.II 
theory (\ref{dual2;theoryA}) and the dual A$^\prime$.II theory 
(\ref{dual2;theoryAA}). 
On the other hand, if the alternative method of deconfining 
(\ref{expanded;theoryB}) is used to expand the symmetric tensor 
$\bS$ and quarks $\bQF$ in anti-fundamentals of the dual B.II theory, 
the dual A$^\prime$.II theory is found to be the dual prescription. 
Similarly, the dual B$^\prime$.II theory can be connected to the 
dual A.II theory under which the expanded method B 
(\ref{expanded;theoryB}) is applied to expand. 
Therefore, there are two types of duality mapping among the four 
dual theories (see Figure 1). 
Duality A derived from the expanded method A (\ref{expanded;theoryA}) 
maps one theory to another theory with different field contents and 
transformation properties. 
Duality B obtained by the expanded method B (\ref{expanded;theoryB}) 
acts as charge conjugation of the $SU(F+5)$ gauge group, 
{\it i.e.} the fields in definite $SU(F+5)$ group representations of 
one theory are mapped to those in complex conjugated representations 
of the other theory.

In order to check the duality, we first find the mapping of the gauge 
invariant chiral operators of the electric theory 
(\ref{electric;theory}) to the magnetic theories. 
The operators are mapped to those of the dual A.II theory in the 
following ways: 
\( M \to M \), \( H \to H \), \( \bar{B} \to (\QS \PbF)^{F+4} Q \), 
\( \Bk \to (\bQF \bQF)^{F-k} (\bS)^{k+5} \) 
(\( k \le {\rm min} (N,F) \)), 
and \( \Vkn \to (\bQF \QS)^{F-k} (\PbF)^{F+4+k+2n} (\WSO)^{2-n} \) 
(\( k \le {\rm min} (N,F) \) and \( n =0,1,2 \)). 
\( \WSO \) is the field strength superfield of the $SO(2F+8)$ 
gauge group of dual A.II theory. 
The mapping of the gauge invariant operators for the 
dual B$^{\prime}$.II is similar to those listed above. 
These gauge invariant operators are mapped to those of 
the dual B.II theory as such: \( M \to M \), \( H \to H \), 
\( \bar{B} \to (\QS \PbF)^{F+4} Q \), 
\( \Bk \to (\bQF \bQF)^{k} (\bS)^{F-k+5} \), 
and \( \Vkn \to (\PF)^{F-k} (\PbF)^{F+4+k+2n} (\WSO)^{2-n} \). 
\( \WSO \) is the field strength superfield of the $SO(2F+8)$ 
gauge group of dual B.II theory. 
There is a similar mapping of the operators for the dual 
A$^{\prime}$.II theory.

Second, we consider the duality by giving an expectation value to 
$\bB$. 
The electric theory is completely higgsed with the remaining 
massless being $\frac{N(N+1)}{2}$ singlets $S$, $NF$ singlets $\QF$, 
and $N(F+4)$ singlets $\bQbF$. 
In the dual A.II theory, 
\( \bB = \langle (\QS \PbF)^{F+4} Q \rangle \ne 0 \) implies 
that the \( SU(F+5) \times SO(2F+8) \) gauge theory is completely 
higgsed. 
The superpotential (\ref{superpotential;dual2A}) generates some 
massive fields. 
After the massive fields being integrated out, only some components 
of $\PbF$, $M$, and $H$ remain massless that exactly match 
the remaining singlet fields $\bQbF$, $\QF$, and $S$ of 
the electric theory. 
Similarly, the \( SU(F+5) \times SO(2F+8) \) gauge group of 
the dual B.II theory is completely higgsed because of 
\( \bB = \langle (\QS \PbF)^{F+4} Q \rangle \ne 0 \) . 
The superpotential (\ref{superpotential;dual2B}) leaves only some 
components of $\PbF$, $M$, and $H$ massless that correspond to 
the massless fields of the electric theory. 
After the massive fields being integrated out, there is no 
superpotential in both dual A.II theory and dual B.II theory 
as in the electric theory.

Next, we analyze the flat directions of the electric theory 
in which $B_k \ne 0$ and $H$ gets an expectation value of rank $r$. 
As discussed above, the electric theory is higgsed to 
\( SO(N-k-r) \) with $\bF+F-k-r$ massless fields in fundamental 
remained. 
In the dual A.II theory, \( \Bk = \langle (\bQF \bQF)^{F-k} 
(\bS)^{k+5} \rangle \ne 0 \) and $H$ having expectation value break 
the gauge symmetry \( SU(F+5) \times SO(2F+8) \) to 
\( SO(k+5) \times SO(2F+8) \) with an 
\( SU(k) \times SU(F-k) \times SU(\bF-r) \) flavor symmetry. 
$M$ is decomposed into $M_a$ with \( (F-k) \times (\bF-r) \) 
components and $M_b$ with \( k \times (\bF-r) \) components. 
In this theory, the $SO(k+5)$ gauge theory has $k+1$ flavors 
in fundamental, and is in confining phase with no dynamical 
generation of superpotential. 
After integrating out massive fields, we find an \( SO(2F+8) \) 
gauge theory with the following matter contents
\begin{equation}
\label{dualityA;deformed}
\matrix{ 
& SO(2F+8) & SU(k) & SU(F-k) & SU(\bF-r)  \cr
\QS  & \Box & {\bf 1} &~\bBox & {\bf 1}  \cr
\PbF  & \Box & {\bf 1} & {\bf 1} &~\bBox  \cr 
M_a  & {\bf 1} & {\bf 1} & \Box & \Box  \cr 
M_b  & {\bf 1} & \Box & {\bf 1} & \Box  \cr
\bS  & {\bf 1} & {\bf 1} & \iiBox & {\bf 1}  \cr
H  & {\bf 1} & {\bf 1} & {\bf 1} & \iiBox  \cr
\bQF^2  & {\bf 1} &~\iibBox & {\bf 1} & {\bf 1} 
}
\end{equation}
with the superpotential \( W = M_a \PbF \QS + \bS \QS^2 + H \PbF^2 + 
(\bQF^2) M_b^2 \PbF^2 \). 
The last term in this superpotential is irrelevant in the infrared, 
and thus can be neglected. 
Rewriting \( \hat{P} = ( \QS, \PbF ) \) and \( \hat{M} = 
\left( \begin{array}{cc} 
\bS & M_a \\ M_a^T & H \end{array} \right) \), we learn that the 
theory (\ref{dualityA;deformed}) is the dual of the higgsed 
electric theory under the $SO(N)$ duality of \cite{IS;so}. 
Now, we consider the same effect to the dual B.II theory. 
\( \Bk = \langle (\bQF \bQF)^{k} (\bS)^{F-k+5} \rangle \ne 0 \) and 
$H$ getting expectation value imply that the 
\( SO(F+5) \times SO(2F+8) \) gauge symmetry is higgsed to 
\( SO(F-k+5) \times SO(2F+8) \) with an 
\( SU(k) \times SU(F-k) \times SU(\bF-r) \) flavor symmetry. 
In this theory, the $SO(F-k+5)$ gauge interaction has $F-k+1$ flavors 
in fundamental, thus is in confining phase without superpotential. 
The result of integrating out massive fields is an $SO(2F+8)$ 
gauge theory with $\bF+F-k-r$ fields in fundamental,
\begin{equation}
\label{dualityB;deformed}
\matrix{ 
& SO(2F+8) & SU(k) & SU(F-k) & SU(\bF-r)  \cr
\PF  & \Box & {\bf 1} &~\bBox & {\bf 1}  \cr
\PbF  & \Box & {\bf 1} & {\bf 1} &~\bBox  \cr 
M_a  & {\bf 1} & {\bf 1} & \Box & \Box  \cr 
M_b  & {\bf 1} & \Box & {\bf 1} & \Box  \cr
\bQF^2  & {\bf 1} & {\bf 1} & \iiBox & {\bf 1}  \cr
H  & {\bf 1} & {\bf 1} & {\bf 1} & \iiBox  \cr
\bS  & {\bf 1} &~\iibBox & {\bf 1} & {\bf 1}
}
\end{equation}
with the superpotential \( W = M_a \PbF \PF + (\bQF^2) \PF^2 
+ H \PbF^2 + \bS M_b^2 \PbF^2 \). 
The last term in the superpotential is negligible in the infrared. 
After rewriting \( \hat{P} = ( \PF, \PbF ) \) and \( \hat{M} = 
\left( \begin{array}{cc} \bQF^2 & M_a \\ M_a^T & H \end{array} 
\right) \), we see that the theory (\ref{dualityB;deformed}) is 
dual under the $SO(N)$ duality of \cite{IS;so}.

Finally, we study the duality of the magnetic theories by adding 
$\lambda S \bQbF \bQbF$ $( \lambda = \lambda^T )$ to the 
superpotential. 
For convenience, we assume that 
\( \lambda = {\rm diag}( \lambda_1, \cdots, \lambda_k, 0, \cdots, 0 ) 
\), \( \QF = \bQbF = 0 \), and \( S \propto {\bf 1} \) in the 
electric theory. 
By deforming it along the flat direction, the electric theory flows 
to $SO(N)$ SUSY gauge theory with $\bF+F-k$ flavors in fundamental. 
The effect of adding $\lambda S \bQbF \bQbF$ to the superpotential 
of the dual B.II theory forces some fields to acquire expectation 
values that break the gauge group $SO(2F+8)$ to $SO(2F-k+8)$. 
Expanding the dual B.II theory around the vacuum and integrating out 
massive fields, we find that the dual B.II theory reduces to an 
\( SO(F+5) \times SO(2F-k+8) \times SU(F) \times SU(\bF-k) \) theory 
in which the $SO(F+5)$ gauge theory confines. 
After integrating out massive fields, we find that the resulting 
theory \( SO(2F-k+8) \times SU(\bF+F-k) \) flows to the dual of 
the deformed electric theory. 
A similar mechanism for the dual A.II theory was discussed 
in \cite{SO}. 

It is interesting to apply this new technique of deconfining to study 
the $SU(N)$ gauge theory with an anti-symmetric tensor. 
This will be done in the forthcoming paper \cite{SU}.

\vspace{1cm}

We would like to thank D. Chang, C. L. Chou, C. Q. Geng, and C. R. Lee 
for useful discussions. 
This work is supported by NSC grant \# 86-2112-M009-034T and a grant 
from Tsing-Hua university.

%%%%%%%%%%%%%%%%%%%%%%%%%%%%%%%%%%%%%%%%%
%%%%%%%%%%%%%%%%%%%%%%%%%%%%%%%%%%%%%%%%%
%%%%%%%%%%%%%%%    Reference   %%%%%%%%%%%%%%%%% 
%%%%%%%%%%%%%%%%%%%%%%%%%%%%%%%%%%%%%%%%%
%%%%%%%%%%%%%%%%%%%%%%%%%%%%%%%%%%%%%%%%%
%%%%%%%%%%%%%%%%%%%%%%%%%%%%%%%%%%%%%%%%%
%%%%%%%%%%%%%%%%%%%%%%%%%%%%%%%%%%%%%%%%%

\end{document}